\documentclass[sigconf, nonacm]{acmart}
\AtBeginDocument{%
  }

\usepackage{graphicx}
\usepackage{subcaption}
\usepackage{enumitem}
\usepackage{multirow}
\usepackage{url}
\usepackage{hyperref}
\usepackage{tabularx}
\usepackage{tcolorbox}

\usepackage{listings}
\usepackage{xcolor}

\definecolor{defblue}{RGB}{30,90,160}
\definecolor{rulegreen}{RGB}{20,120,70}
\definecolor{rulepurple}{RGB}{110,70,150}

\definecolor{promptbg}{RGB}{245,245,245}
\definecolor{promptframe}{RGB}{160,160,160}

\newtcolorbox{promptbox}{
  colback=promptbg,
  colframe=promptframe,
  boxrule=0.4pt,
  arc=1mm,
  left=0.5mm,
  right=0.5mm,
  top=0.5mm,
  bottom=0.5mm,
  boxsep=0.5mm
}
\newboolean{showcomments}
\setboolean{showcomments}{false}
\ifthenelse{\boolean{showcomments}}
  {% If true: Define the visible comment style (NEW STYLE, NO BOX)
    \newcommand{\nb}[2]{%
      {\small\bfseries\sffamily\textcolor{green}{#1}}%
      ~{\sf\small$\blacktriangleright$\textit{\textcolor{red}{#2}}$\blacktriangleleft$}%
    }
  }
  {% If false: Define commands to do nothing
    \newcommand{\nb}[2]{}
    
  }

\newcommand\jie[1]{\nb{Jie}{#1}}

\setcopyright{none}
\renewcommand\footnotetextcopyrightpermission[1]{}
\begin{document}

%%
%% The "title" command has an optional parameter,
%% allowing the author to define a "short title" to be used in page headers.
\title{Benchmarking and Evaluating VLMs for Software Architecture Diagram Understanding}

\author{Shuyin Ouyang}
% \email{shuyin.ouyang@kcl.ac.uk}
\affiliation{%
  \institution{King's College London}
  \city{London}
  \country{United Kingdom}
}

\author{Jie M. Zhang}
% \email{jie.zhang@kcl.ac.uk}
\affiliation{%
  \institution{King's College London}
  \city{London}
  \country{United Kingdom}
}

\author{Jingzhi Gong}
% \email{jingzhi.gong@kcl.ac.uk}
\affiliation{%
  \institution{King's College London}
  \city{London}
  \country{United Kingdom}
}

\author{Gunel Jahangirova}
% \email{gunel.jahangirova@kcl.ac.uk}
\affiliation{%
  \institution{King's College London}
  \city{London}
  \country{United Kingdom}
}

\author{Mohammad Reza Mousavi}
% \email{mohammad.mousavi@kcl.ac.uk}
\affiliation{%
  \institution{King's College London}
  \city{London}
  \country{United Kingdom}
}

\author{Jack Johns}
% \email{jack.johns@bt.com}
\affiliation{%
  \institution{British Telecom}
  \city{London}
  \country{United Kingdom}
}

\author{Beum Seuk Lee}
% \email{rom.lee@bt.com}
\affiliation{%
  \institution{British Telecom}
  \city{London}
  \country{United Kingdom}
}

\author{Adam Ziolkowski}
% \email{adam.ziolkowski@bt.com}
\affiliation{%
  \institution{British Telecom}
  \city{London}
  \country{United Kingdom}
}

\author{Botond Virginas}
% \email{botond.virginas@bt.com}
\affiliation{%
  \institution{British Telecom}
  \city{London}
  \country{United Kingdom}
}

\author{Joost Noppen}
% \email{johannes.noppen@bt.com}
\affiliation{%
  \institution{British Telecom}
  \city{London}
  \country{United Kingdom}
}

\begin{abstract}
% Recent progress in large language models has substantially advanced code-centric software engineering tasks such as code generation, testing, and maintenance. However, early-stage design artifacts remain understudied.
% In particular, software architecture diagrams play a central role in communicating system structure, behavioral, and data relationships, yet the ability of modern vision-language models (VLMs) to understand such diagrams remains underexplored.
Software architecture diagrams are important design artifacts for communicating system structure, behavior, and data organization throughout the software development lifecycle.
Although recent progress in large language models has substantially advanced code-centric software engineering tasks such as code generation, testing, and maintenance, the ability of modern vision-language models (VLMs) to understand software architecture diagrams remains underexplored.
To address this gap, we present SADU, a benchmark for \textbf{S}oftware \textbf{A}rchitecture \textbf{D}iagram \textbf{U}nderstanding that evaluates VLMs on architecture diagrams as structured software engineering artifacts rather than generic images. 
SADU contains 154 carefully curated diagrams spanning behavioral, structural, and ER diagrams, paired with structured annotations and 2,431 question-answer tasks covering counting and retrieval reasoning. 
We evaluate 11 state-of-the-art VLMs from the Gemini, Claude, GPT, and Qwen families.

Our results show that software architecture diagram understanding remains challenging for current models: the best-performing model gemini-3-flash-preview achieves only 70.18\% accuracy, while gpt-4o-mini only achieves 17.77\% accuracy.
The results further reveal the weaknesses in diagram reasoning and visual relation grounding, highlighting a gap between current VLMs and the needs of design-stage software engineering. 
SADU provides a foundation for future research on diagram-aware AI systems and more faithful AI-assisted software engineering workflows.
\end{abstract}

\maketitle

\section{Introduction}

Large language models (LLMs) have recently achieved strong performance in many software engineering tasks, especially those focused on source code, such as code generation~\cite{chen2021evaluating, ouyang2025dscodebench}, code testing~\cite{wang2025testeval}, and code maintenance~\cite{jimenez2023swe}.
As a result, the benchmark ecosystem in software engineering has rapidly matured around code-centric stages of the Software Development Life Cycle (SDLC)~\cite{SDLC}.
However, software engineering is not only about producing correct code.
It also depends on early-stage artifacts that define what should be built and how the system should be organized before implementation begins.
Compared with downstream coding tasks, these design-oriented capabilities remain much less studied.

Among design artifacts, software architecture diagrams are particularly important.
They encode components, interfaces, dependencies, interaction flows, deployment structures, and data relationships in a compact visual form, and are widely used to communicate system design across stakeholders~\cite{li2024devbench}.
In practice, engineers rely on such diagrams to validate assumptions, align on high-level constraints, and guide implementation decisions.
A model that cannot correctly interpret architecture diagrams is less likely to reliably support upstream software engineering activities that depend on diagram understanding, and may introduce inconsistencies between design intent and downstream development~\cite{hou2024vision, sun2025math}.

Recent vision-language models (VLMs) provide a promising foundation for automating document-centric software engineering workflows, since architecture diagrams are visual artifacts tightly coupled with technical semantics~\cite{wang2025document, nacson2025docvlm, li2024devbench}.
If VLMs could robustly understand software architecture diagrams, they could support tasks such as extracting design knowledge, answering architecture-level questions, and assisting architecture-aware development.
Despite this potential, current evaluation protocols provide limited visibility into these capabilities~\cite{hou2024vision}.
Existing multi-modal benchmarks are typically general-purpose and not tailored to software engineering semantics, while mainstream software engineering benchmarks largely ignore architecture diagrams~\cite{jimenez2023swe}.

To address this gap, we present \textbf{SADU}, a benchmark for \underline{\textbf{S}}oftware \underline{\textbf{A}}rchitecture \underline{\textbf{D}}iagram \underline{\textbf{U}}nderstanding.
% SADU evaluates VLMs on architecture diagrams as structured software engineering artifacts rather than generic images.
% We construct SADU by collecting real-world software architecture diagrams from diverse sources, manually curating them for quality and coverage, and annotating each diagram with structured representations and question-answer pairs.
We construct SADU by collecting real-world software architecture diagrams, including behavioral, structural, and entity--relationship (ER) diagrams from diverse sources \jie{did you report quality and coverage? Otherwise it is risky} and annotating each diagram with structured representations and question-answer pairs.
The benchmark includes diverse diagram families commonly used in practice and pairs each diagram with structured question-answer tasks that probe both recognition and reasoning. 
% Using SADU, we evaluate 11 state-of-the-art VLMs and show that software architecture diagram understanding remains challenging: even the best-performing model achieves only 70.18\% overall accuracy\jie{duplicate with the following}.
% These results reveal substantial weaknesses in diagram reasoning and visual relation grounding, and highlight the need for diagram-aware AI systems for software engineering.

Our experiments show that software architecture diagram understanding remains challenging for current VLMs.
Across the 11 evaluated models, the best overall accuracy is only 70.18\%, achieved by gemini-3-flash-preview, followed by claude-sonnet-4.5 at 56.36\% and gpt-5-nano at 55.45\%.
We further observe clear variation across diagrams: 
ER diagrams are generally the easiest, whereas behavioral diagrams are more difficult and exhibit larger performance gaps across models, suggesting that reasoning over interaction flows and component behaviors is more challenging than recognizing more explicit structural patterns.
In addition, stronger model capability does not necessarily translate into better cost efficiency.
Although some reasoning-enabled models generate substantially longer outputs, the performance gains are not always proportional to the additional token cost.
In contrast, gemini-3.1-flash-lite-preview achieves a comparatively favorable balance between accuracy and efficiency.
% \jie{we want to first highlight that overall existing VLMs need significant improvement in diagram understanding}
% Overall, current VLMs still require substantial improvement in software architecture diagram understanding.
These results reveal a substantial capability gap in existing VLMs on understanding software architecture diagrams in both effectiveness and efficiency.
% Overall, these results reveal a substantial capability gap in existing VLMs on software architecture diagrams, especially for reasoning over behavioral interactions, and highlight the importance of evaluating both effectiveness and efficiency in this setting.

To summarize, this paper makes the following contributions:
\begin{itemize}[leftmargin=*]
    \item We introduce SADU, a benchmark for evaluating VLMs' ability to understand software architecture diagrams.
    \item We conduct a comprehensive empirical study of 11 state-of-the-art VLMs on SADU, together with the impact of different prompting strategies.
    % \jie{do not highlight the two assessment in the introduction as it's trivial contribution}
    % rule-based evaluation and LLM-as-a-judge assessment.
    \item We provide detailed analyses of model failure modes and 
    offer actionable insights for researchers and developers.
    % \item \jie{merge this with the previous one}We study the impact of different prompting strategies on software architecture diagram understanding.
\end{itemize}

The remainder of this paper is organized as follows.
Section~\ref{section: SADU} introduces SADU.
Section~\ref{section: Experiment Design} presents the experimental design, including the research questions, evaluated models, and measurement metrics.
Section~\ref{section: Result} reports the empirical results and key findings.
Section~\ref{section: Discussion} discusses case studies, limitations, and implications.
Section~\ref{section: Related Work} reviews related work.
Section~\ref{section: Conclusion} concludes the paper.
Section~\ref{section: Data Availability Statement} shows the data availability statement.

\section{SADU Benchmark}
\label{section: SADU}

SADU is designed to evaluate VLMs' ability to interpret, ground, and reason over software architecture diagrams.
The benchmark consists of 154 diagrams spanning three representative diagram families: 51 behavioral diagrams, 53 structural diagrams, and 50 ER diagrams.
% Behavioral diagrams describe how a system acts over time, focusing on dynamic interactions, control flow, and message exchange among components or actors. 
% Structural diagrams represent the static organization of a system, showing its components, modules, classes, or services and the relationships among them.
% ER diagrams describe entities, their attributes, and the relationships among entities in a database or information system.
% These categories capture complementary aspects of architecture understanding, including runtime interactions (behavioral), static component organization (structural), and data-centric conceptual modeling (ER).
We introduce the three types of diagrams as follows:
% \jie{say here that in the following we will explain A,B, and C}

\paragraph{Behavioral Diagram.}
A behavioral diagram illustrates the \emph{dynamic behavior} of a system, focusing on how components interact over time.
It is used to represent workflows, message passing, control flow, or service interactions during system operation.
In software architecture, behavioral diagrams typically emphasize runtime processes, such as how a request moves across services or how different modules collaborate to complete a task.

\paragraph{Structural Diagram.}
A structural diagram illustrates the \emph{static organization} of a system, focusing on what components exist and how they are related.
It is used to represent architectural elements such as classes, modules, subsystems, or components, together with their connections, dependencies, inheritance, or containment relationships.
In software architecture, structural diagrams emphasize the overall system organization rather than its runtime behavior.

\paragraph{ER Diagram.}
An Entity--Relationship (ER) diagram illustrates the \emph{data structure} of a system, focusing on entities, their attributes, and the relationships among them.
It is commonly used in database and information system design to describe how data objects are organized and associated.
In software architecture, ER diagrams emphasize the schema-level view of the system, such as which entities exist, what properties they contain, and how they are linked.

\begin{figure}[h!]
\centering
\includegraphics[width=\linewidth]{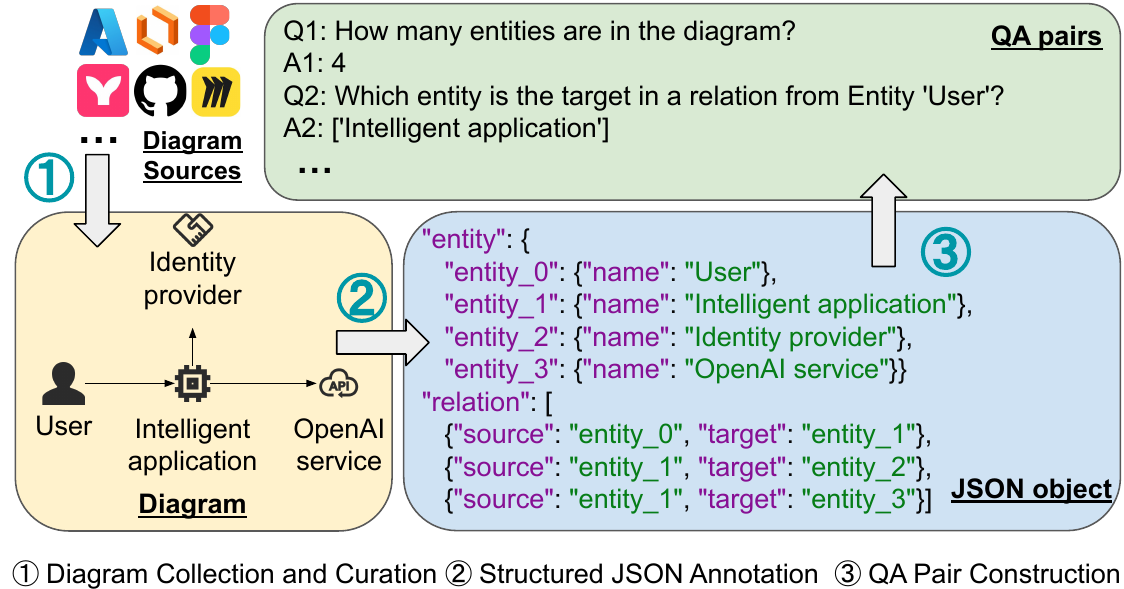}
\vspace*{-0.4cm}
\caption{Overview of the SADU construction pipeline.}
\label{fig: SADU construction}
\vspace*{-0.4cm}
\end{figure}

\begin{table}[h!]
\caption{Benchmark statistics.}
\vspace*{-0.3cm}
\centering
\resizebox{\linewidth}{!}{%
\begin{tabular}{l rr r r r r r r r}
\toprule
\multirow{3}{*}{Diagram Type} & \multicolumn{6}{c}{JSON object}  & \multicolumn{2}{c}{QA pairs} \\
& \multicolumn{3}{c}{Entity} & \multicolumn{3}{c}{Relation} & & \\
\cmidrule{2-9}
 & Max & Min & Mean  & Max  & Min  & Mean & QA pairs & subtypes \\
\midrule
Behavioral & 37 & 4 & 12.20 & 26 & 2 & 8.75 & 810 & 20 \\
Structural & 29 & 3 & 15.85 & 44 & 2 & 19.91 & 848 & 20 \\
ER & 16 & 2 & 7.60 & 22 & 1 & 7.72 & 773 & 18 \\
\midrule
Total & 37 & 2 & 11.96 & 44 & 1 & 12.25 & 2431 & 24 \\

\bottomrule
\end{tabular}
}
\vspace*{-0.4cm}
\label{table: benchmark statistics}
\end{table}

\paragraph{\textbf{Diagram Collection and Curation}}

% \shuyin{TODO: enrich the following}

We construct SADU from a combination of real-world sources.
Behavioral diagrams are collected from Microsoft Azure Architecture documentation~\cite{azure}.
Structural diagrams are obtained from the PyUNML dataset~\cite{PyUNML}.
ER diagrams are collected from public demos on diagram-design platforms, including Mermaid~\cite{mermaid}, Lucidchart~\cite{lucidchart}, Figma~\cite{figma}, and Miro~\cite{miro}.

We then apply a curation process to improve quality.
Specifically, we filter out instances with low resolution or unreadable text, and retain only diagrams with sufficient visual clarity for evaluation.
We also consider diagram complexity during selection to preserve diversity across categories, including the number of entities, relation density, and the presence or absence of clusters.
Starting from an initial pool of 1,044 collected diagrams, we first remove 318 duplicated diagrams, 157 low-resolution images, and 64 diagrams with unreadable text, leaving 465 candidates; we then further filter out 311 diagrams based on relevance to software architecture and annotation quality, resulting in a final benchmark of 154 diagrams.

\paragraph{\textbf{Manual JSON Annotation}}

To make the benchmark task-oriented and model-agnostic, we define a unified set of key diagram elements across all diagram types.
The element definitions are shown in Figure~\ref{fig: VLM prompt}.
These definitions are used consistently in annotation, question construction, and evaluation to reduce ambiguity and improve comparability across diagram families.

Each diagram is paired with a JSON metadata file that provides a machine-readable representation of its content.
For every diagram, we record:
(1) a canonical list of entities with unique identifiers and textual labels;
(2) the set of relations between entities, including directionality and optional relation labels; and
(3) the set of clusters, as well as optional per-entity attributes and methods when present in the diagram.
This structured representation (i) provides a ground-truth abstraction for consistent QA construction; and
(ii) enables deterministic evaluation without relying on subjective visual interpretation during scoring.
The annotation process required two authors and approximately 160 human hours of manual effort.
% \jie{details missing: how many people did the annotation?}.

\paragraph{\textbf{QA Pair Construction}}

For each diagram, we construct a set of QA pairs that probe different levels of diagram understanding, ranging from counting tasks (e.g., the number of entities or relation labels) to retrieval tasks (e.g., identifying the correct elements that satisfy a query).
Each diagram is associated with a QA JSON file that stores the natural language questions and corresponding answers, where the questions are automatically generated from subtype templates.
The templates can be found on our homepage~\cite{homepage}.

In total, SADU includes 24 QA subtypes (shown in Figure~\ref{fig: error analysis (subtype)}) covering a broad range of capabilities, including element retrieval (e.g., entities, relations, and clusters), counting and comparison (e.g., the number of incoming/outgoing edges and the largest cluster), and relational reasoning (e.g., identifying the source or target of a given entity).
To reduce ambiguity, we manually curate the questions and retain only those with a unique, well-defined answer. 
In the end, SADU comprises 2,431 QA pairs.
% \jie{details missing for the manual process}

% For each diagram, we construct a set of QA pairs that probe different levels of diagram understanding, ranging from counting problems (e.g., figuring out how many entities and relation labels) to retrieval problems (e.g., identifying the correct elements based on queries).
% Each diagram is associated with a QA JSON file that stores the natural language question and its answer.
% In total, SADU includes 24 QA subtypes, covering a broad range of skills, such as element retrieval (entities, relations, clusters), counting and comparison (e.g., number of incoming/outgoing edges, largest cluster), and relational reasoning (the source or target of a certain entity).
% To remove the ambiguity of the question, we curate the questions and select the questions that have one and only one answer.

\section{Experiment Design}
\label{section: Experiment Design}

This section describes the experimental design, including the research questions, evaluated VLMs, and measurement metrics.

\subsection{Research Questions}

Our experiments are designed to systematically evaluate VLMs on software architecture diagram understanding.
We organize the study around the following research questions:

\noindent \textbf{RQ1: How well do state-of-the-art VLMs perform on SADU?}
This question examines the overall effectiveness of current VLMs on software architecture diagram understanding.

\noindent \textbf{RQ2: What are the main failure modes of VLMs on SADU?}
This question investigates why VLMs fail on certain questions through fine-grained error analysis.
Specifically, we study:

RQ2.1: Error analysis on counting questions.

RQ2.2: Error analysis on retrieval questions.

\noindent \textbf{RQ3: How do different prompting strategies affect model performance?}
This question investigates how prompt design influences performance and failure behavior.

\noindent \textbf{RQ4: How do reasoning configurations affect diagram understanding performance?}
This question investigates whether different reasoning settings influence VLM performance on software architecture diagram understanding.

\noindent \textbf{RQ5: How consistent are rule-based evaluation and LLM-as-a-judge assessment?}
This question evaluates the reliability of LLM-as-a-judge and its agreement with rule-based scoring.

\subsection{VLMs}

We evaluate models from four VLM families: Gemini, Claude, GPT, and Qwen, covering both proprietary and open-weight systems as well as a range of model sizes.

\textbf{Gemini:} gemini-3-flash-preview~\cite{gemini-3-flash-preview}, gemini-3.1-flash-lite-preview~\cite{gemini-3.1-flash-lite-preview}, gemini-2.5-flash~\cite{gemini-2.5-flash}, and gemini-2.5-flash-lite~\cite{gemini-2.5-flash-lite}.

\textbf{Claude:} claude-haiku-4.5~\cite{claude-haiku-4-5} and claude-sonnet-4.5~\cite{claude-sonnet-4-5}.

\textbf{GPT:} gpt-5-nano~\cite{gpt-5-nano} and gpt-4o-mini~\cite{gpt-4o-mini}.

\textbf{Qwen:} qwen-2.5-VL~\cite{qwen2.5-vl} in 32B, 7B, and 3B variants.

We use the prompt template shown in Figure~\ref{fig: VLM prompt} to construct inputs.
To reduce randomness, we set the temperature of all models to 0.
Additional implementation details, including model-specific parameters and prompt variants, are provided on our homepage~\cite{homepage}.

\begin{figure}[h]
\centering
\begin{minipage}{0.95\columnwidth}
\begin{promptbox}
\small
Please answer the question strictly based on the given diagram and follow ALL rules below.

{\color{defblue}
Definitions:

1. Entity: An individual, identifiable component in the diagram, represented by a distinct icon or label with text.

2. Cluster: A visual container or boundary (circle, box, or similar shape) that encloses multiple related entities to indicate environment, region, or service domain.

3. Relation: Any directional, bidirectional, or non-directional interaction between entities shown in the diagram.

4. Relation label: The exact string attached to the relation arrow or line.

5. Attribute: A descriptive element attached to an entity in the diagram, shown as text inside or near the entity, representing properties, configurations, or characteristics of that entity.

6. Method: An operational element associated with an entity in the diagram, shown as a named action, capability, or functional label that indicates what behavior the entity performs or exposes.
}

{\color{rulegreen}
Answering Rules:

1. Count all visible items shown inside the diagram.

2. Name matching is case sensitive.

3. Do NOT include explanations, reasoning, or text outside the JSON object.

4. Output MUST follow the exact format shown below, including the [start] and [end] markers, the key name "answer", array brackets, and quotes around every string.

5. Do NOT add extra fields, extra text, or comments.

6. The JSON array MUST contain only the final answer values (e.g., entity names, cluster names, relation labels, counts, etc., depending on the question).
}

{\color{rulepurple}
How many entities are in the diagram?
}

Required Output Format (strict).

[start]

\{"answer": x\}

[end]

The answer MUST be a valid JSON object.

\end{promptbox}
\end{minipage}
\vspace*{-0.3cm}
\caption{Example prompt used for requesting VLMs. The prompt includes definitions, rules, and questions. 
% \jie{what is the question in the prompt? Is it an example, or is this just the prompt for that question? }
}
\label{fig: VLM prompt}
\vspace*{-0.4cm}
\end{figure}

\subsection{Measurement}

We evaluate VLMs on SADU along two axes: effectiveness and cost.
Effectiveness is measured using two complementary evaluation protocols:
(i) an \textbf{LLM-as-a-judge} evaluator that provides a more flexible semantic assessment of prediction correctness, and
(ii) a \textbf{rule-based} scorer that provides deterministic matching against the gold answers while offering a more lightweight and sustainable evaluation alternative.
% (ii) a \textbf{rule-based} scorer that provides deterministic matching against the gold answers.
Cost is measured by the \textbf{token usage} of each request, including prompt, completion, and reasoning tokens when available.

\subsubsection{LLM-as-a-Judge Evaluation}

We employ an LLM-as-a-judge evaluator based on gpt-5.4~\cite{gpt-5.4} to assess prediction correctness under a more flexible semantic interpretation.
The judge is given the question, the diagram, the gold answer, and the model prediction, and returns a binary decision indicating whether the prediction should be considered correct.
We use judge-based accuracy to quantify robustness to surface-form variation and to identify cases where strict rule-based matching may underestimate true performance.
Due to space limitations, the judge prompt template is provided on our project homepage~\cite{homepage}.

\subsubsection{Rule-based Evaluation} Unless otherwise specified, we use exact match (EM) to determine whether a model answers a question correctly.
Given an instance $i$ with gold answer $G_i$ and model prediction $P_i$, the rule-based EM indicator is:
\begin{equation}
\mathrm{EM}_i =
\begin{cases}
1 & \text{if } P_i = G_i,\\
0 & \text{otherwise.}
\end{cases}
\end{equation}
Overall accuracy is $\mathrm{Acc}=\frac{1}{N}\sum_{i=1}^{N}\mathrm{EM}_i$, where $N$ is the number of QA instances.
\paragraph{Token Cost.}
We report token usage per request as:
\begin{equation}
\mathrm{Token}_i = \mathrm{Token}_{\mathrm{Prompt},i} + \mathrm{Token}_{\mathrm{Completion},i} + \mathrm{Token}_{\mathrm{Reason},i},
\end{equation}
and aggregate it by the mean over the evaluation set.

\textbf{Counting Questions.} Counting questions require predicting a scalar value (e.g., the number of entities or relations).
In addition to strict EM, we use three complementary measurements.

\paragraph{Tolerance Accuracy (Acc@$\pm k$).}
Because counting errors are often near-miss mistakes (e.g., off-by-one due to small perceptual misses), we report tolerance accuracy to capture approximate correctness:
\begin{equation}
\mathrm{Acc@}\pm k = \frac{1}{N}\sum_{i=1}^{N}\mathbb{I}\left(\left|\hat{y}_i - y_i\right|\le k\right),
\end{equation}
where $y_i$ is the gold count and $\hat{y}_i$ is the predicted count.
We use $k\in\{1,2\}$ to quantify how often predictions are approximately correct even when they are not exact matches.

\paragraph{Direction Bias.}
To test whether models systematically over-count or under-count, we measure signed error:
\begin{equation}
\mathrm{Bias} = \frac{1}{N}\sum_{i=1}^{N}\left(\hat{y}_i - y_i\right).
\end{equation}
A positive (negative) value indicates a tendency to predict larger (smaller) counts than the ground truth.

\paragraph{Mean Absolute Error (MAE)}
We report MAE to quantify the average deviation magnitude:
\begin{equation}
\mathrm{MAE} = \frac{1}{N}\sum_{i=1}^{N}\left|\hat{y}_i - y_i\right|.
\end{equation}

\textbf{Retrieval Questions.}
Retrieval questions require returning a set of elements (e.g., entity names or relation labels).
Let $G_i$ be the gold set and $P_i$ be the predicted set for instance $i$.

\paragraph{Precision, Recall, and F1.}
\begin{equation}
\mathrm{Precision}_i = \frac{|P_i \cap G_i|}{|P_i|+\epsilon}, \quad
\mathrm{Recall}_i  = \frac{|P_i \cap G_i|}{|G_i|+\epsilon},
\end{equation}
\begin{equation}
\mathrm{F1}_i = \frac{2\cdot \mathrm{Precision}_i\cdot \mathrm{Recall}_i}{\mathrm{Precision}_i+\mathrm{Recall}_i+\epsilon},
\end{equation}
where $\epsilon$ is a small constant to avoid division by zero.
We report macro-averaged scores across instances.

\paragraph{Exact-Match, Subset, and Superset Rates.}
We further characterize set-level prediction behavior via:
\begin{equation}
\mathrm{EMRate} = \frac{1}{N}\sum_{i=1}^{N}\mathbb{I}(P_i = G_i),
\end{equation}
\begin{equation}
\mathrm{SubsetRate} = \frac{1}{N}\sum_{i=1}^{N}\mathbb{I}(P_i \subsetneq G_i),
\quad
\mathrm{SupersetRate} = \frac{1}{N}\sum_{i=1}^{N}\mathbb{I}(P_i \supsetneq G_i).
\end{equation}
SubsetRate captures under-retrieval behavior, while SupersetRate captures over-retrieval behavior.

\paragraph{Missing and Spurious Elements.}
To make retrieval errors more interpretable, we report the average numbers of missing and spurious elements:
\begin{equation}
\mathrm{Missing}_i = |G_i \setminus P_i|,
\quad
\mathrm{Spurious}_i = |P_i \setminus G_i|,
\end{equation}
\begin{equation}
\overline{\mathrm{Missing}} = \frac{1}{N}\sum_{i=1}^{N}\mathrm{Missing}_i,
\quad
\overline{\mathrm{Spurious}} = \frac{1}{N}\sum_{i=1}^{N}\mathrm{Spurious}_i.
\end{equation}

% We introduce the following metrics for measuring the performance of VLMs on our benchmark.

% We evaluate VLMs on SADU along two axes: \emph{effectiveness} and \emph{efficiency}.
% Effectiveness is measured by the correctness of predicted answers under two complementary evaluation protocols:
% (i) a \textbf{rule-based} scorer that deterministically compares model outputs against gold answers, and
% (ii) an \textbf{LLM-as-a-judge} scorer that adjudicates borderline cases (e.g., minor surface-form variations) and provides a robustness check for rule-based matching.
% Efficiency is measured by the \textbf{token cost} of each request, including prompt tokens and completion tokens.

% \textbf{Cost}: 

\section{Results and Findings}
\label{section: Result}

This section shows the experiment results and findings.

\subsection{RQ1: How well do state-of-the-art VLMs perform on SADU?}

\textbf{Overall Model Performance.}
Table~\ref{table: accuracy} reports the accuracy of VLMs across the three types of software architecture diagrams.
Among all evaluated models, gemini-3-flash-preview achieves the best overall accuracy (70.18\%), followed closely by gemini-2.5-flash (69.68\%) and gemini-3.1-flash-lite-preview (66.31\%).
Overall, Gemini models achieve the strongest performance on SADU, which is consistent with Gemini's strong multimodal reasoning ability in its technique report~\cite{comanici2025gemini}.
Claude models are competitive but consistently below the top Gemini models.
Specifically, claude-sonnet-4.5 reaches 56.36\%, while claude-haiku-4.5 achieves 48.25\%, suggesting a substantial performance gap between the two variants.
Among OpenAI models, gpt-5-nano reaches 55.45\%, which is comparable to claude-sonnet-4.5 but at a substantially higher inference cost, whereas gpt-4o-mini performs much worse at 17.77\%.
Open-source Qwen models trail the closed-source models, although they exhibit a clear scaling trend: qwen-2.5-VL-32B reaches 45.17\%, compared with 38.58\% for qwen-2.5-VL-7B and 31.30\% for qwen-2.5-VL-3B.

\textbf{Performance Across Diagram Types.}
Model performance varies noticeably across diagram families.
% behavioral
Behavioral diagrams exhibit the largest performance variance across models.
While the strongest Gemini models remain above 59\%, smaller models, such as qwen-2.5-VL-3B, drop to 29.26\%.
% structural
Structural diagrams show moderate difficulty overall.
Many models obtain results similar to those on behavioral diagrams, although some models exhibit noticeable drops.
For instance, claude-sonnet-4.5 performs worse on structural diagrams (46.70\%) than on behavioral diagrams (53.83\%).
% ER
ER diagrams are generally the easiest, with most models achieving their highest accuracy on this category.
For example, gemini-2.5-flash reaches 82.54\% on ER diagrams, substantially higher than its performance on behavioral (62.10\%) and structural (65.21\%) diagrams.
% This suggests that ER diagrams expose more explicit entity--relationship patterns that are easier for VLMs to recognize.
% This pattern suggests that reasoning over behavioral flows and interactions between components is more challenging than recognizing more explicit structural patterns.

% \subsubsection{Error Analysis by Problem Subtypes}

% \shuyin{TODO: move this to RQ1 }

\begin{figure}[h!]
\centering
  \begin{subfigure}{\linewidth}
    % \centering
    \includegraphics[width=1.0\linewidth]{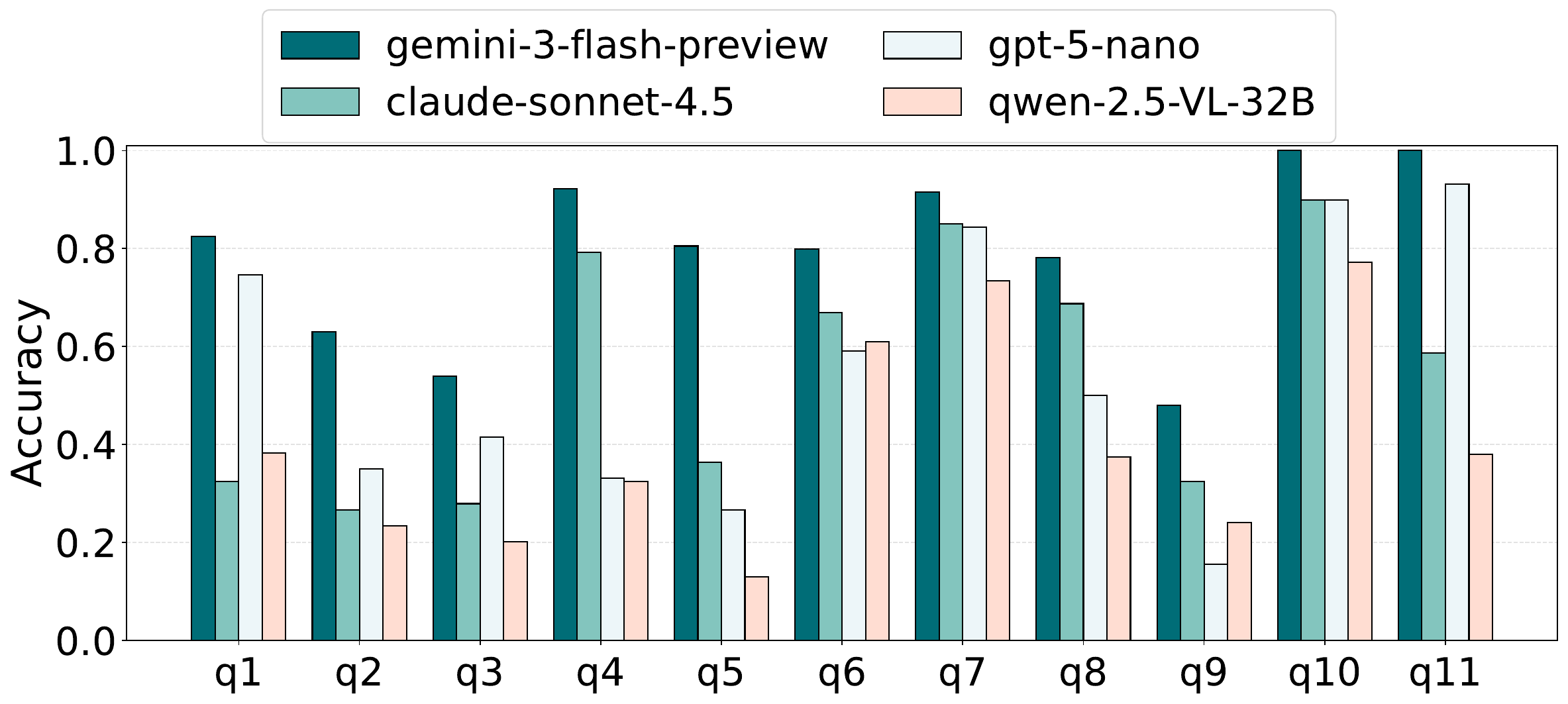}
    \caption{Counting questions. 
q1: total entity number;
q2: total relation number;
q3: entities that have more than one relation;
q4: total cluster number;
q5: how many entities are not in clusters;
q6: how many relations have label;
q7: how many clusters have clusters within;
q8: how many entities a certain cluster has;
q9: how many relation types;
q10: how many attributes certain entity has;
q11: how many methods certain entity has.}
  \end{subfigure}
    \begin{subfigure}{\linewidth}
    % \centering
    \includegraphics[width=1.0\linewidth]{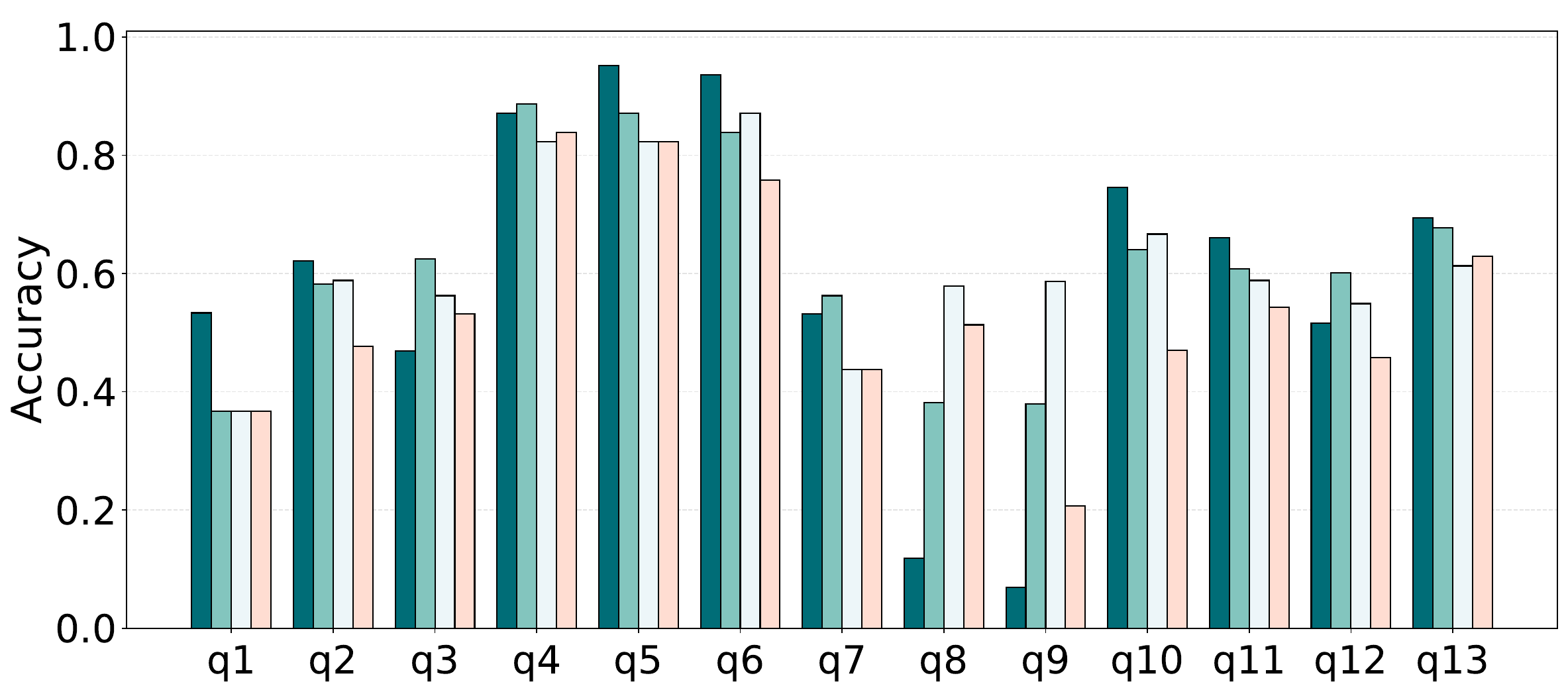}
    \caption{Retrieval questions.
q1: which cluster has the most entities;
q2: relation statement;
q3: which entity a certain cluster contains;
q4: label on the relation;
q5: label on the relation source;
q6: label on the relation target;
q7: cluster contains certain entity;
q8: which entity contains a certain attribute;
q9: which entity contains a certain method;
q10: type on the relation;
q11: relation source;
q12: relation target;
q13: which relation has a certain label.}
  \end{subfigure}
% \vspace*{-0.3cm}
\caption{RQ1: Performance across question subtypes.}
\vspace*{-0.4cm}
\label{fig: error analysis (subtype)}
\end{figure}

\begin{figure}[h!]
\centering
% \vspace*{-0.2cm}
\includegraphics[width=0.9\linewidth, trim=0 0 0 0.26cm, clip]{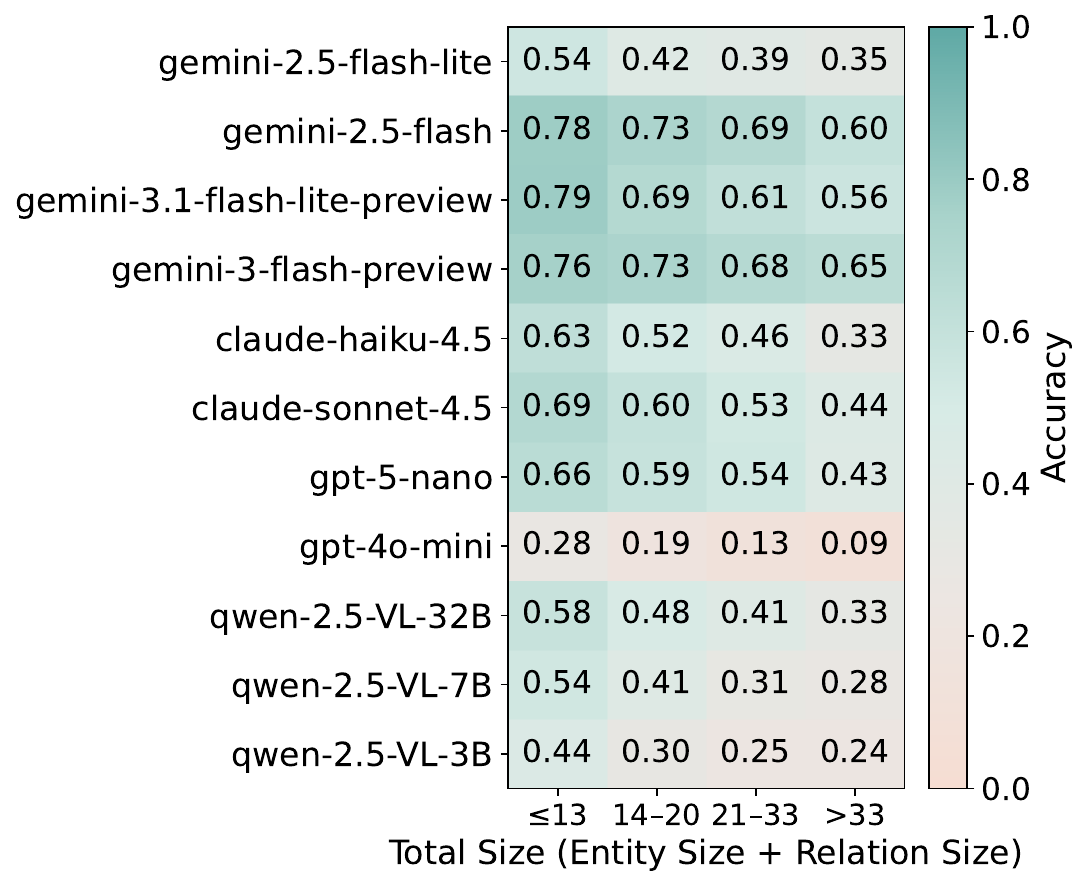}
\vspace*{-0.3cm}
\caption{RQ1: Model accuracy across diagram complexity.}
\label{fig: complexity}
\vspace*{-0.3cm}
\end{figure}

\textbf{Performance Across Question Subtypes.} 
% To further understand which question types are most challenging for current VLMs, we analyze performance across individual problem subtypes.
Figure~\ref{fig: error analysis (subtype)} reports subtype-level accuracy for four representative models, covering both counting and retrieval questions.
Across the counting subtypes in Figure~\ref{fig: error analysis (subtype)}(a), the strongest models maintain relatively high accuracy on most tasks, and several counting questions are nearly saturated.
In particular, q10 and q11 are answered perfectly by gemini-3-flash-preview and gpt-5-nano.
Similarly, q4 and q7 also achieve high accuracy for the stronger models.
These subtypes require counting visually explicit and locally grouped elements, which current VLMs can handle relatively well.
The hardest counting questions include q8 and q9, where all models show noticeably lower accuracy.
% Questions like q3 are also more challenging than simple counting tasks.
% These results suggest that counting becomes difficult when the model must aggregate information across multiple entities or relations rather than count a directly visible structure.

Retrieval questions exhibit a larger spread in difficulty, as shown in Figure~\ref{fig: error analysis (subtype)}(b).
Some relation-label retrieval subtypes are relatively easy.
For example, q4, q5, and q6 achieve high accuracy across the stronger models, with gemini-3-flash-preview and claude-sonnet-4.5 performing particularly well.
These tasks depend on identifying visually explicit labels attached to relation arrows, which appears to be comparatively accessible to current VLMs.
However, retrieval tasks that require precise grounding to entities or relations remain much more difficult.
In particular, q8 and q9 are among the hardest subtypes, especially for gemini-3-flash-preview, gpt-5-nano, and qwen-2.5-VL-32B.
Likewise, q13 is consistently harder than direct relation-label reading tasks.
These patterns indicate that models struggle more when they must associate a textual attribute, method, or label with the correct structural element and then return that element, rather than simply read or count visually explicit content.
We also conduct a case study in Section~\ref{subsec: Case Study} on several hard diagram patterns to further analyze the failures.

\textbf{Performance Over Complexity.} 
Figure~\ref{fig: complexity} shows a clear negative correlation between diagram complexity and model accuracy: performance declines steadily as the total number of entities and relations increases. 
This trend holds across nearly all evaluated models, confirming that larger and denser diagrams impose a substantially greater challenge for current VLMs. 
Among the tested VLMs, the Gemini family remains the most robust under increasing complexity.
In particular, \texttt{gemini-3-flash-preview} and \texttt{gemini-2.5-flash} maintain relatively strong performance even on the largest diagrams, whereas weaker models such as \texttt{gpt-4o-mini} and the smaller Qwen variants show much sharper degradation. 
These results suggest that diagram complexity is a key factor affecting VLM performance on SADU.

% Figure~\ref{fig: complexity} shows that accuracy consistently decreases as diagram size increases, indicating that larger diagrams with more entities and relations are substantially harder for all models. 
% Among the evaluated systems, the Gemini models remain the most robust under increasing complexity, while weaker models such as gpt-4o-mini and the smaller Qwen variants exhibit much sharper performance degradation.

% \shuyin{TODO: analyze the heatmap}

% \begin{tcolorbox}
% \textbf{\underline{Finding 6:}}
% Subtype-level analysis reveals a clear difficulty gradient: counting and explicit relation-label reading are comparatively easy, whereas cross-element aggregation and precise grounding of attributes, methods, and labeled relations remain challenging for current VLMs.
% \end{tcolorbox}

\textbf{Cost Efficiency.}
Table~\ref{table: token usage} reports the mean completion token usage of different models.
Most models generate relatively concise responses, typically within 13--20 tokens, including gemini-2.5-flash-lite, gemini-3.1-flash-lite-preview, gpt-4o-mini, and the qwen-2.5-VL variants.
In contrast, more advanced models with built-in reasoning capabilities produce longer outputs.
For example, gemini-2.5-flash and gemini-3-flash-preview generate more than 300 completion tokens on average, while gpt-5-nano produces 1475.19 completion tokens on average, nearly two orders of magnitude higher than most other evaluated models.
Despite higher token usage, gpt-5-nano does not outperform the strongest Gemini models in accuracy, suggesting a less favorable efficiency profile.

\begin{tcolorbox}
\textbf{\underline{Answer to RQ1:}}
State-of-the-art VLMs achieve only moderate performance on SADU, with the best model, gemini-3-flash-preview, reaching 70.18\% overall accuracy.
Gemini-3.1-flash-lite-preview offers a strong accuracy--cost trade-off, achieving 66.31\% overall accuracy with only 16.94 average completion tokens.
\end{tcolorbox}

% \begin{tcolorbox}
% \textbf{\underline{Answer to RQ1:}}
% VLM performance varies substantially across diagram families.
% ER diagrams are generally the easiest, whereas behavioral diagrams are more challenging and exhibit larger performance gaps across models.
% \end{tcolorbox}

% \begin{tcolorbox}
% \textbf{\underline{Answer 1.2:}}
% Among the evaluated models, gemini-3.1-flash-lite-preview provides a favorable balance between performance and cost, achieving relatively high accuracy while maintaining very low token consumption.
% \end{tcolorbox}

\begin{table}[h!]
\caption{Accuracy for different VLMs. Gemini-3-flash-preview performs the best among all the VLMs.}
\vspace*{-0.2cm}
\centering
\resizebox{\columnwidth}{!}{%
\begin{tabular}{l r r r r}
\toprule

Model & Behavioral & Structural & ER & All \\
\midrule

gemini-2.5-flash-lite & 42.59\% & 36.44\% & 50.71\% & 43.03\% \\
gemini-2.5-flash & 62.10\% & 65.21\% & 82.54\% & 69.68\% \\
gemini-3.1-flash-lite-preview & 59.51\% & 60.85\% & 79.43\% & 66.31\% \\
\textbf{gemini-3-flash-preview} & \textbf{63.58\%} & \textbf{68.87\%} & \textbf{78.53\%} & \textbf{70.18\%} \\
claude-haiku-4.5 & 48.27\% & 38.56\% & 58.86\% & 48.25\% \\
claude-sonnet-4.5 & 53.83\% & 46.70\% & 69.60\% & 56.36\% \\
gpt-5-nano & 48.64\% & 50.59\% & 67.92\% & 55.45\% \\
gpt-4o-mini & 19.14\% & 10.14\% & 24.71\% & 17.77\% \\
qwen-2.5-VL-32B & 44.44\% & 36.44\% & 55.50\% & 45.17\% \\
qwen-2.5-VL-7B & 34.20\% & 31.01\% & 51.49\% & 38.58\% \\
qwen-2.5-VL-3B & 29.26\% & 26.77\% & 38.42\% & 31.30\% \\

\bottomrule
\end{tabular}
}
\vspace*{-0.4cm}
\label{table: accuracy}
\end{table}

\begin{table}[h!]
\caption{RQ2.1: Mean completion token usage. Models with reasoning ability (gemini-2.5-flash, gemini-3-flash-preview, and gpt-5-nano) cost more tokens.}
\vspace*{-0.2cm}
\centering
\resizebox{0.9\columnwidth}{!}{%
\begin{tabular}{l r r r r}
\toprule

Model & Behavioral & Structural & ER & All \\
\midrule

gemini-2.5-flash-lite & 16.47 & 18.62 & 16.40 & 17.20 \\
gemini-2.5-flash & 339.33 & 324.37 & 289.35 & 318.22 \\
gemini-3.1-flash-lite-preview & 16.37 & 18.03 & 16.35 & 16.94 \\
gemini-3-flash-preview & 455.17 & 450.54 & 439.39 & 448.54 \\
claude-haiku-4.5 & 34.69 & 50.99 & 39.50 & 41.90 \\
claude-sonnet-4.5 & 22.76 & 43.81 & 32.88 & 33.32 \\
gpt-5-nano & 1536.71 & 1641.64 & 1228.12 & 1475.19 \\
gpt-4o-mini & 14.76 & 14.71 & 13.06 & 14.20 \\
qwen-2.5-VL-32B & 12.96 & 14.28 & 13.09 & 13.46 \\
qwen-2.5-VL-7B & 13.23 & 14.37 & 13.01 & 13.56 \\
qwen-2.5-VL-3B & 16.93 & 16.31 & 19.53 & 17.54 \\
\bottomrule
\end{tabular}
}
\vspace*{-0.4cm}
\label{table: token usage}
\end{table}

\subsection{RQ2: What are the main failure modes of VLMs on SADU?}

\subsubsection{Error Analysis on Counting Questions}

To better understand why VLMs fail on counting questions, we conduct a fine-grained error analysis using three complementary metrics: MAE, tolerance accuracy (Acc@$\pm k$), and direction bias.
Counting questions are challenging because they require both accurate visual perception and reliable reasoning over diagram elements.

Table~\ref{table: counting MAE} reports the MAE of different models.
Among all evaluated systems, gemini-3-flash-preview achieves the lowest overall MAE (0.54), followed by gemini-2.5-flash (0.85) and gemini-3.1-flash-lite-preview (1.12), indicating that these models generally predict counts close to the ground truth.
In contrast, weaker models exhibit substantially larger deviations.
For example, gpt-4o-mini has an overall MAE of 3.80, while qwen-2.5-VL-3B reaches 3.97, suggesting that these models often miscount multiple diagram elements.

% \begin{tcolorbox}
% \textbf{\underline{Finding 3:}}
% % Counting performance shows a clear quality threshold across models:
% Stronger VLMs keep the overall MAE close to or below 1, whereas weaker models often exceed an MAE of 3. 
% This gap suggests that counting errors do not degrade gradually, but instead increase sharply once models lack sufficiently reliable visual grounding and counting ability.
% \end{tcolorbox}

\begin{table}[h!]
\caption{MAE for counting problems}
\vspace*{-0.2cm}
\centering
\resizebox{0.9\linewidth}{!}{
\begin{tabular}{l r r r r}
\toprule

Model & Behavioral & Structural & ER & All \\
\midrule
gemini-2.5-flash-lite & 2.36 & 4.46 & 2.43 & 3.12 \\
gemini-2.5-flash & 1.10 & 1.03 & 0.42 & 0.85 \\
gemini-3.1-flash-lite-preview & 1.33 & 1.50 & 0.50 & 1.12 \\
\textbf{gemini-3-flash-preview} & \textbf{0.85} & \textbf{0.49} & \textbf{0.30} & \textbf{0.54} \\
claude-haiku-4.5 & 1.79 & 3.29 & 1.49 & 2.20 \\
claude-sonnet-4.5 & 1.46 & 2.38 & 0.96 & 1.62 \\
gpt-5-nano & 1.80 & 2.93 & 1.94 & 2.24 \\
gpt-4o-mini & 2.99 & 5.83 & 2.43 & 3.80 \\
qwen-2.5-VL-32B & 1.98 & 3.44 & 2.24 & 2.58 \\
qwen-2.5-VL-7B & 2.57 & 4.15 & 2.20 & 3.00 \\
qwen-2.5-VL-3B & 3.60 & 5.03 & 0.78 & 3.97 \\
\bottomrule
\end{tabular}
}\vspace*{-0.3cm}
\label{table: counting MAE}

\end{table}

To further explore counting failures, Table~\ref{table: tolerance accuracy (counting)} reports tolerance accuracy under relaxed thresholds (Acc@$\pm1$ and Acc@$\pm2$).
The results show that many incorrect predictions are still close to the correct value, indicating that counting failures are often small mistakes rather than completely incorrect estimates.
Given that the average numbers of entities and relations are 11.96 and 12.25, respectively, deviations within $\pm1$ or $\pm2$ can reasonably be regarded as near-miss predictions.
For example, gemini-2.5-flash achieves an exact-match accuracy of 0.72 overall, which increases to 0.84 under Acc@$\pm1$ and 0.91 under Acc@$\pm2$.
Similarly, gemini-3-flash-preview improves from 0.76 exact-match accuracy to 0.88 and 0.92 under the relaxed thresholds.
In contrast, weaker models benefit much less from tolerance-based evaluation.
For instance, qwen-2.5-VL-3B achieves only 0.08 exact-match accuracy overall, which increases only slightly to 0.11 under Acc@$\pm1$.
This suggests that its counting predictions are often not merely off by one or two, but substantially different from the ground truth.

% \begin{tcolorbox}
% \textbf{\underline{Finding 4:}}
% A large fraction of counting errors are near misses, especially for stronger models, suggesting that many failures arise from off-by-one or off-by-two mistakes.
% \end{tcolorbox}

Finally, we analyze the direction of counting errors using the bias metric reported in Table~\ref{table: directional bias (counting)}.
Positive values indicate overestimation, while negative values indicate underestimation.
Several weaker models exhibit a clear under-counting tendency, particularly on structural diagrams.
For example, gpt-4o-mini shows a strong negative bias (-3.24) on structural diagrams, indicating that it frequently predicts substantially smaller counts than the ground truth.
Similarly, qwen-2.5-VL-3B exhibits a bias of -2.54 on structural diagrams, suggesting that it often fails to identify all relevant elements.
In contrast, stronger models show smaller biases and more balanced counting behavior.
For example, gemini-3-flash-preview remains close to zero across most diagram categories, indicating relatively well-calibrated predictions, while gemini-2.5-flash shows only a mild positive overall bias (0.40).

Overall, these results suggest that counting errors are closely related to failures in reliably identifying all relevant diagram elements.
The prevalence of near-miss errors among stronger models indicates that many failures are small local counting mistakes, whereas the strong under-counting tendency of weaker models suggests more substantial perceptual omissions.
Improving performance on these tasks will likely require better visual grounding and more robust object-level reasoning over diagram structure.

\begin{tcolorbox}
\textbf{\underline{Answer to RQ2.1:}}
% Stronger VLMs keep the overall MAE close to or below 1, whereas weaker models often exceed an MAE of 3. 
Around 70\% Acc@$\pm2$ shows that most counting errors are near misses, especially for stronger models, suggesting that many failures arise from off-by-one or off-by-two mistakes.
Stronger VLMs keep the MAE close to or below 1, whereas weaker models often exceed an MAE of 3, which suggests that counting errors do not degrade gradually, but instead increase sharply once models lack sufficiently reliable visual counting ability.
\end{tcolorbox}

\begin{table*}[h!]
\caption{RQ2.1: Error analysis for counting questions. We use $Acc$ for exact match, $\mathrm{Acc@}\pm k$ for tolerance accuracy.}
\vspace*{-0.2cm}
\centering
\resizebox{0.8\linewidth}{!}{%
\begin{tabular}{l | r r r | r r r | r r r | r r r}
\toprule

\multirow{2}{*}{Model} & \multicolumn{3}{|c|}{Behavioral} & \multicolumn{3}{c|}{Structural} & \multicolumn{3}{c|}{ER} & \multicolumn{3}{c}{All} \\
\cmidrule{2-13}
 & Acc & Acc@$\pm$1 & Acc@$\pm$2 & Acc & Acc@$\pm$1 & Acc@$\pm$2 & Acc & Acc@$\pm$1 & Acc@$\pm$2 & Acc & Acc@$\pm$1 & Acc@$\pm$2\\
\midrule
gemini-2.5-flash-lite & 0.26 & 0.49 & 0.63 & 0.31 & 0.43 & 0.49 & 0.38 & 0.58 & 0.67 & 0.32 & 0.49 & 0.59 \\
gemini-2.5-flash & 0.56 & 0.74 & 0.88 & 0.73 & 0.85 & 0.90 & 0.86 & 0.92 & 0.95 & 0.72 & 0.84 & 0.91 \\
gemini-3.1-flash-lite-preview & 0.49 & 0.74 & 0.87 & 0.59 & 0.78 & 0.85 & 0.75 & 0.93 & 0.96 & 0.61 & 0.81 & 0.89 \\
gemini-3-flash-preview & 0.60 & 0.80 & 0.90 & 0.81 & 0.89 & 0.92 & 0.86 & 0.94 & 0.94 & 0.76 & 0.88 & 0.92 \\
claude-haiku-4.5 & 0.36 & 0.61 & 0.77 & 0.39 & 0.57 & 0.64 & 0.56 & 0.73 & 0.81 & 0.44 & 0.63 & 0.74 \\
claude-sonnet-4.5 & 0.43 & 0.68 & 0.83 & 0.46 & 0.65 & 0.74 & 0.66 & 0.85 & 0.91 & 0.51 & 0.73 & 0.82 \\
gpt-5-nano & 0.38 & 0.59 & 0.76 & 0.52 & 0.65 & 0.71 & 0.59 & 0.70 & 0.74 & 0.50 & 0.65 & 0.74 \\
gpt-4o-mini & 0.16 & 0.39 & 0.57 & 0.09 & 0.19 & 0.35 & 0.24 & 0.41 & 0.56 & 0.16 & 0.33 & 0.49 \\
qwen-2.5-VL-32B & 0.30 & 0.62 & 0.77 & 0.39 & 0.53 & 0.62 & 0.45 & 0.63 & 0.72 & 0.38 & 0.59 & 0.70 \\
qwen-2.5-VL-7B & 0.22 & 0.47 & 0.65 & 0.29 & 0.41 & 0.49 & 0.40 & 0.60 & 0.68 & 0.31 & 0.49 & 0.61 \\
qwen-2.5-VL-3B & 0.03 & 0.06 & 0.08 & 0.12 & 0.16 & 0.17 & 0.09 & 0.11 & 0.12 & 0.08 & 0.11 & 0.12 \\
\bottomrule
\end{tabular} 
}
\vspace*{-0.2cm}
\label{table: tolerance accuracy (counting)}
\end{table*}

\begin{table*}[h!]
\caption{RQ2.1: Error analysis for counting questions. We use direction bias to test whether models systematically over-count or under-count. Positive values indicate a tendency to overestimate counts, while negative values indicate underestimation.}
\vspace*{-0.2cm}
\centering
\resizebox{0.85\linewidth}{!}{%
\begin{tabular}{l | r r r | r r r | r r r | r r r}
\toprule

% Diagram Type & Exact Match & Acc@$\pm$1 & Acc@$\pm$2 \\
\multirow{2}{*}{Model} & \multicolumn{3}{|c|}{Behavioral} & \multicolumn{3}{c|}{Structural} & \multicolumn{3}{c|}{ER} & \multicolumn{3}{c}{All} \\
\cmidrule{2-13}
 & Bias & P(Bias>0) & P(Bias<0) & Bias & P(Bias>0) & P(Bias<0) & Bias & P(Bias>0) & P(rBias<0) & Bias & P(Bias>0) & P(Bias<0)\\
\midrule
gemini-2.5-flash-lite & -0.40 & 0.34 & 0.36 & -0.91 & 0.25 & 0.42 & 0.21 & 0.28 & 0.33 & -0.38 & 0.29 & 0.37 \\
gemini-2.5-flash & 0.30 & 0.28 & 0.15 & 0.66 & 0.17 & 0.10 & 0.32 & 0.10 & 0.04 & 0.43 & 0.18 & 0.10 \\
gemini-3.1-flash-lite-preview & -0.01 & 0.29 & 0.22 & 0.07 & 0.10 & 0.31 & 0.20 & 0.11 & 0.14 & 0.09 & 0.17 & 0.22 \\
gemini-3-flash-preview & 0.15 & 0.28 & 0.10 & 0.03 & 0.06 & 0.10 & 0.22 & 0.08 & 0.04 & 0.13 & 0.13 & 0.08 \\
claude-haiku-4.5 & -0.16 & 0.31 & 0.30 & -0.52 & 0.27 & 0.29 & 0.15 & 0.29 & 0.14 & -0.18 & 0.29 & 0.25 \\
claude-sonnet-4.5 & -0.30 & 0.25 & 0.33 & -0.04 & 0.23 & 0.30 & 0.01 & 0.18 & 0.16 & -0.11 & 0.22 & 0.26 \\
gpt-5-nano & 0.16 & 0.36 & 0.26 & -0.59 & 0.22 & 0.26 & 0.44 & 0.25 & 0.16 & -0.02 & 0.27 & 0.23 \\
gpt-4o-mini & -1.04 & 0.39 & 0.45 & -3.23 & 0.43 & 0.48 & 0.03 & 0.41 & 0.35 & -1.46 & 0.41 & 0.43 \\
qwen-2.5-VL-32B & -0.69 & 0.33 & 0.37 & -1.06 & 0.24 & 0.37 & 0.06 & 0.29 & 0.26 & -0.57 & 0.28 & 0.34 \\
qwen-2.5-VL-7B & -0.10 & 0.42 & 0.36 & 0.17 & 0.35 & 0.36 & 1.07 & 0.32 & 0.28 & 0.38 & 0.36 & 0.33 \\
qwen-2.5-VL-3B & -1.01 & 0.07 & 0.08 & -2.54 & 0.12 & 0.18 & 0.17 & 0.03 & 0.02 & -1.71 & 0.08 & 0.10 \\
\bottomrule
\end{tabular}
}
\vspace*{-0.4cm}
\label{table: directional bias (counting)}
\end{table*}

\subsubsection{Error Analysis on Retrieval Problems}

% In addition to counting questions, SADU includes retrieval problems, which require models to return a set of diagram elements, such as entities or relations, that satisfy a given condition.
% To analyze model behavioral on these tasks, we use several complementary metrics, including Precision, Recall, F1, exact-match rate (EM Rate), subset and superset rates, and the average numbers of missing and spurious elements.
Compared with counting tasks, retrieval problems are more demanding because models must identify, ground, and return multiple relevant elements as a complete set.
Table~\ref{table: error analysis (retrieval)} reports retrieval performance across different VLMs.
Among all evaluated models, gemini-3-flash-preview achieves the highest F1 score (0.79), followed by gemini-2.5-flash (0.78) and gemini-3.1-flash-lite-preview (0.77).
These models also achieve relatively high EM rates, indicating a stronger ability to return the correct element set exactly.
In contrast, weaker models perform substantially worse.
For example, gpt-4o-mini achieves an F1 score of only 0.38 with an EM rate of 0.18, while qwen-2.5-VL-3B reaches an F1 score of 0.53 but only a 0.26 EM rate, suggesting that these models often retrieve partially correct answers.

To better understand retrieval failures, we analyze subset and superset behaviors.
The subset rate measures cases where the predicted set is a strict subset of the ground truth.
For example, gemini-3-flash-preview shows a relatively high subset rate (0.21), despite achieving the highest overall F1 score.
Combined with its high precision (0.87) and low superset rate (0.01), this suggests that strong models can even avoid spurious elements but sometimes omit a small number of correct ones.
This interpretation is consistent with the average number of missing elements, which ranges from 1.72 to 2.39.
The superset rate measures cases where the predicted set contains additional incorrect elements.
Several weaker models exhibit notable over-retrieval behavior.
For example, gpt-4o-mini has a relatively high superset rate (0.19), suggesting that it frequently introduces extra incorrect elements when retrieving diagram components.
This tendency is also reflected in the average number of spurious elements, which remains high for several weaker models, including gpt-4o-mini (3.57) and qwen-2.5-VL-3B (3.90).

\begin{tcolorbox}
\textbf{\underline{Answer to RQ2.2:}}
Retrieval errors primarily reflect two recurring failure modes: missing relevant diagram elements and introducing spurious ones.
Even strong models often return incomplete sets, while weaker models are more prone to over-retrieval and noisy predictions.
\end{tcolorbox}

\begin{table*}[h!]
\caption{RQ2.2: Error analysis for retrieval problems}
\vspace*{-0.3cm}
\centering
\resizebox{0.8\linewidth}{!}{%
\begin{tabular}{l r r r r r r r r}
\toprule

Model & Precision & Recall & F1 & EMRate & SubsetRate & SupersetRate & $\overline{\mathrm{Missing}}$ & $\overline{\mathrm{Spurious}}$ \\
\midrule
gemini-2.5-flash-lite & 0.70 & 0.69 & 0.68 & 0.53 & 0.09 & 0.06 & 1.79 & 5.85 \\
gemini-2.5-flash & 0.79 & 0.79 & 0.78 & 0.63 & 0.10 & 0.07 & 1.85 & 2.71 \\
gemini-3.1-flash-lite-preview & 0.79 & 0.77 & 0.77 & 0.68 & 0.08 & 0.02 & 1.81 & 2.27 \\
gemini-3-flash-preview & 0.87 & 0.76 & 0.79 & 0.58 & 0.21 & 0.01 & 2.39 & 1.53 \\
claude-haiku-4.5 & 0.67 & 0.66 & 0.65 & 0.49 & 0.10 & 0.06 & 1.84 & 4.05 \\
claude-sonnet-4.5 & 0.73 & 0.73 & 0.72 & 0.58 & 0.08 & 0.07 & 1.82 & 4.14 \\
gpt-5-nano & 0.72 & 0.68 & 0.69 & 0.57 & 0.10 & 0.03 & 1.90 & 2.98 \\
gpt-4o-mini & 0.36 & 0.46 & 0.38 & 0.18 & 0.03 & 0.19 & 2.06 & 3.57 \\
qwen-2.5-VL-32B & 0.72 & 0.66 & 0.68 & 0.53 & 0.15 & 0.02 & 1.86 & 1.71 \\
qwen-2.5-VL-7B & 0.65 & 0.60 & 0.61 & 0.46 & 0.12 & 0.04 & 1.72 & 2.87 \\
qwen-2.5-VL-3B & 0.59 & 0.52 & 0.53 & 0.26 & 0.10 & 0.01 & 1.96 & 3.90 \\

\bottomrule
\end{tabular}
}
\vspace*{-0.4cm}
\label{table: error analysis (retrieval)}
\end{table*}

\subsection{RQ3: How do different prompting strategies affect model performance?}

To investigate the impact of prompt design on software architecture diagram understanding, we evaluate four prompt variants using Gemini models: \textit{Full prompt}, \textit{Without definition}, \textit{Without rules}, and \textit{Only question}.
We focus on Gemini models in this analysis because they achieve the strongest overall performance in our main evaluation, and restricting the study to this model family keeps the cost manageable.
The full prompt includes both domain-specific definitions and explicit answering rules, while the other variants progressively remove these components.
Table~\ref{table: different prompt} reports the results.

The full prompt achieves strong and stable performance across all evaluated models, indicating that providing both task definitions and answer-format constraints helps models better interpret diagram elements and produce valid outputs.
For example, gemini-2.5-flash achieves 67.91\% overall accuracy with the full prompt, while gemini-3-flash-preview reaches 68.00\%.
Removing domain definitions leads to a noticeable drop in performance for most models.
For instance, the overall accuracy of gemini-2.5-flash-lite decreases from 40.89\% to 34.27\%, and gemini-3.1-flash-lite-preview drops from 63.84\% to 57.88\%.
This suggests that explicit definitions help models interpret diagram concepts such as entities, clusters, and relations, especially when the visual structure is ambiguous or complex.

In contrast, removing the answering rules has a smaller effect on overall accuracy and even leads to slight improvements for some models.
For example, gemini-2.5-flash increases marginally from 67.91\% to 68.04\%, and gemini-3-flash-preview rises from 68.00\% to 68.61\%.
This result suggests that explicit answer-format constraints are less important than domain definitions for overall task performance, although they may still help reduce formatting errors and improve output consistency.
When only the question is provided, performance drops substantially for most models.
For example, gemini-2.5-flash decreases from 67.91\% to 49.61\%, and gemini-3.1-flash-lite-preview falls from 63.84\% to 52.90\%.
This indicates that relying on the question alone, without domain definitions or answer-format guidance, makes it significantly harder for models to correctly interpret both diagram elements and task requirements.

\begin{tcolorbox}
\textbf{\underline{Answer to RQ3:}}
The prompting strategy has a substantial impact on diagram understanding performance.
Domain-specific definitions consistently improve accuracy, whereas explicit answering rules have a comparatively smaller effect on overall performance.
\end{tcolorbox}

% Across the different diagram categories, similar trends are observed. behavioral diagrams and Structural diagrams are more sensitive to prompt changes, while ER diagrams remain relatively stable, suggesting that ER diagrams contain more explicit structural information that models can interpret even with minimal prompting.

% The results demonstrate that prompt design plays a critical role in diagram reasoning tasks. 
% Providing clear domain definitions significantly improves model performance, while strict answering rules have limited impact on accuracy. 
% These findings highlight the importance of including domain-specific guidance when designing prompts for software architecture diagram understanding tasks.

\begin{table}[h!]
\caption{RQ3: Effect of different prompts on VLM performance (Gemini only)}
\vspace*{-0.2cm}
\centering
\resizebox{\columnwidth}{!}{%
\begin{tabular}{l l r r r r}
\toprule

Prompt & Model & Behavioral & Structural & ER & All \\
\midrule
\multirow{4}{*}{Full} & gemini-2.5-flash-lite & 42.59\% & 36.44\% & 50.71\% & 43.03\% \\
 & gemini-2.5-flash & 62.10\% & 65.21\% & \textbf{82.54\%} & 69.68\% \\
 & gemini-3.1-flash-lite-preview & 59.51\% & 60.85\% & 79.43\% & 66.31\% \\
 & gemini-3-flash-preview & 63.58\% & 68.87\% & 78.53\% & 70.18\% \\
\midrule
\multirow{4}{*}{Without definition}  & gemini-2.5-flash-lite & 35.80\% & 27.24\% & 45.15\% & 35.79\% \\
 & gemini-2.5-flash & 57.65\% & 56.49\% & 76.07\% & 63.10\% \\
 & gemini-3.1-flash-lite-preview & 53.95\% & 54.60\% & 71.67\% & 59.81\% \\
 & gemini-3-flash-preview & 63.21\% & 68.40\% & 78.91\% & 70.01\% \\
\midrule
\multirow{4}{*}{Without rules}  & gemini-2.5-flash-lite & 46.05\% & 39.86\% & 58.47\% & 47.84\% \\
 & gemini-2.5-flash & 64.69\% & 66.27\% & 81.50\% & 70.59\% \\
 & gemini-3.1-flash-lite-preview & 60.25\% & 60.85\% & 81.63\% & 67.26\% \\
 & gemini-3-flash-preview & \textbf{65.31\%} & \textbf{69.34\%} & 80.47\% & \textbf{71.53\%} \\
\midrule
\multirow{4}{*}{Only question} & gemini-2.5-flash-lite & 36.67\% & 34.08\% & 45.54\% & 38.58\% \\
 & gemini-2.5-flash & 32.22\% & 31.72\% & 41.14\% & 34.88\% \\
 & gemini-3.1-flash-lite-preview & 51.36\% & 60.97\% & 69.47\% & 60.47\% \\
 & gemini-3-flash-preview & 46.05\% & 49.17\% & 52.78\% & 49.28\% \\
 
\bottomrule
\end{tabular}
}
\vspace*{-0.4cm}
\label{table: different prompt}
\end{table}

\subsection{RQ4: How do different thinking levels affect model performance?}

Table~\ref{table: different thinking level} examines how different thinking levels affect the performance of gemini-3-flash-preview under two prompt settings: \textit{full} and \textit{only question}.
We use gemini-3-flash-preview for this analysis because it is the only evaluated model that currently provides a configurable thinking-level option.

Under the \textit{full} prompt setting, the \textit{low} thinking level achieves the best overall performance, reaching 72.85\% accuracy on the \textit{All} category, which is higher than both the \textit{default} setting (70.18\%) and the \textit{high} setting (69.77\%).
This improvement is mainly driven by gains on behavioral and \textit{ER} diagrams, where \textit{low} thinking reaches 64.44\% and 86.55\%, respectively, both of which are the best results in the table.
For Structural diagrams, however, the \textit{high} thinking level performs best at 69.10\%, slightly exceeding \textit{default} (68.87\%) and \textit{low} (68.40\%).
Overall, these results suggest that increasing the thinking level does not consistently improve performance, and that a lower thinking setting may be more effective in this task configuration.

These results suggest that overthinking~\cite{huang2025mitigating, cuadron2025danger} is a major source of error in this setting.
To further investigate its impact, we conduct a deeper analysis of all results between \textit{low} and \textit{high} thinking levels.
Through two authors' manual analysis of the responses, we identify 470 cases exhibiting overthinking-related errors.
% \jie{manual details missing, at least two authors involved? also there is no comparison with low thinking so the usefulness is very limited}
For counting problems (262 cases), overthinking mainly manifests in two forms.
First, \textbf{task confusion} (50.4\%), where the \textit{high} thinking setting returns a list of entity or label strings instead of a number.
Second, \textbf{scope inflation} (23.7\%), where the \textit{high} thinking setting often produces an excessively large count.
This bias is asymmetric, with overcounting occurring much more frequently than undercounting (69 cases vs.\ 24 cases).
For retrieval problems (208 cases), overthinking is also reflected in two major error patterns.
First, \textbf{missing elements} (54.8\%), where the model over-deliberates about which items truly qualify and therefore omits valid answers.
Second, \textbf{wrong substitution} (21.6\%), where the predicted answer contains the correct number of elements but replaces correct items with incorrect ones, often drawn from another part of the same diagram.

\begin{tcolorbox}
\textbf{\underline{Answer to RQ4:}}
\textit{Low} thinking with the \textit{full} prompt achieves the best performance with 72.85\% accuracy.
Increasing the thinking level to \textit{high} does not yield additional gains, while prompt completeness has a much larger effect on performance.
\end{tcolorbox}

% A clearer pattern emerges when comparing prompt settings.
% Across all thinking levels, the \textit{full} prompt substantially outperforms the \textit{only question} prompt.
% For example, under the \textit{default} setting, the overall accuracy drops from 70.18\% to 49.28\% when the model is given only the question.
% Similar gaps are observed for \textit{low} thinking (72.85\% vs.\ 52.24\%) and \textit{high} thinking (69.77\% vs.\ 48.75\%).
% This trend is consistent across all three diagram categories, indicating that providing complete prompt context is much more important than adjusting the thinking level alone.
% In other words, increasing the reasoning budget does not compensate for missing prompt guidance.

% \begin{tcolorbox}
% \textbf{\underline{Finding 8:}}
% The best-performing configuration is \textit{low} thinking with the \textit{full} prompt.
% Increasing the thinking level to \textit{high} does not provide additional gains, while prompt completeness has a much larger effect on performance.
% \end{tcolorbox}

\begin{table}[h!]
\caption{RQ4: Different thinking levels' effect on VLM's performance (Gemini-3-flash-preview only)}
\vspace*{-0.2cm}
\centering
\resizebox{\columnwidth}{!}{%
\begin{tabular}{l l r r r r}
\toprule

Thinking level & Prompt Type & Behavioral & Structural & ER & All \\
\midrule

\multirow{2}{*}{default} & full & 63.58\% & 68.87\% & 78.53\% & 70.18\% \\
 & only question & 46.05\% & 49.17\% & 52.78\% & 49.28\% \\
\multirow{2}{*}{low} & full & \textbf{64.44\%} & 68.40\% & \textbf{86.55\%} & \textbf{72.85\%} \\
 & only question & 46.30\% & 53.30\% & 57.31\% & 52.24\% \\
\multirow{2}{*}{high} & full & 62.10\% & \textbf{69.10\%} & 78.53\% & 69.77\% \\
 & only question & 45.06\% & 48.58\% & 52.78\% & 48.75\% \\
\bottomrule
\end{tabular}
}
\vspace*{-0.4cm}
\label{table: different thinking level}
\end{table}

\subsection{RQ5: How consistent are LLM-as-a-judge and rule-based evaluation?}

Table~\ref{table: different LLM judge and rule-based judge} compares the judgments produced by LLM-as-a-judge and rule-based evaluation.
We treat the LLM-as-a-judge result as the reference and categorize each prediction into four cases: \textit{T\_Same}, \textit{F\_Same}, \textit{T\_Diff}, and \textit{F\_Diff}.
Here, \textit{T\_Same} and \textit{F\_Same} indicate agreement between the two evaluation methods, whereas \textit{T\_Diff} and \textit{F\_Diff} indicate disagreement.
% This comparison helps us assess how closely rule-based evaluation aligns with LLM-based assessment.
Overall, most models show high agreement between the two evaluation methods.
For example, gemini-2.5-flash, gemini-3-flash-preview, claude-haiku-4.5, claude-sonnet-4.5, and qwen-2.5-VL-32B all achieve more than 90\% combined agreement (\textit{T\_Same} + \textit{F\_Same}).
This suggests that, in most cases, the LLM judge reaches the same conclusion as the rule-based evaluator.
In particular, several stronger models, such as gemini-2.5-flash and gemini-3-flash-preview, have large \textit{T\_Same} portions (64.29\% and 63.51\%, respectively), indicating that LLM-as-a-judge is largely consistent with rule-based evaluation when identifying correct answers for these models.

% \jie{following text is chaotic, please rewrite}
To better understand why rule-based evaluation is stricter than LLM-as-a-judge, we collect 3,908 disagreement cases (from 11 models) where the LLM judge labels a prediction as True, but the rule-based evaluator labels it as False (\textit{T\_Diff}).
These cases consist of 2,872 counting questions and 1,036 retrieval questions across all 11 models.
For \textbf{counting questions}, we observe three main patterns.
The largest category is close-value errors (40.5\%), where the predicted count differs from the ground truth by 2 to 5.
This is followed by off-by-one errors (35.9\%), where the prediction is only one count away from the correct answer, and large-error cases (23.6\%), where the prediction differs by more than 5.
For \textbf{retrieval questions}, the disagreement patterns are more revealing.
51.7\% of the cases are completely different answers, where the predicted set has no clear correspondence to the ground truth.
The remaining cases mainly involve partial lexical similarity, including substring match cases (28.9\%), where a predicted element is a prefix or suffix of a gold element, and fuzzy matching cases (18.0\%), where the predicted element differs from a correct one by only a small edit distance.

\begin{tcolorbox}
\textbf{\underline{Answer to RQ5:}}
LLM-as-a-judge and rule-based evaluation show high overall agreement, with rates ranging from 85.73\% to 95.60\% for most models.
Rule-based evaluation is deterministic but can be overly strict for numerically close or lexically similar answers, 
whereas LLM-as-a-judge is more flexible but may also accept answers that are clearly different from ground truth.
\jie{give number: how much rejected? also cannot tell this conclusion from the text above}
\end{tcolorbox}

\begin{table}[h!]
\caption{RQ5: Difference between LLM-as-a-judge and rule-based evaluation. T\_Same/F\_Same means LLM-as-a-judge judges the result as True/False, and rule-based evaluation judges the same. T\_Diff/F\_Diff means LLM-as-a-judge judges the result as True/False, while a rule-based evaluation judges it as False. All\_Same is the sum of T\_Same and F\_Same. 
% \jingzhi{I think we mentioned we can have a aggregated col like "agreement rate", so it is more obvious they are highly aligned}
}
\vspace*{-0.2cm}
\centering
\resizebox{\columnwidth}{!}{%
\begin{tabular}{l r r r r r}
\toprule

Model & All\_Same & T\_Same & F\_Same & T\_Diff & F\_Diff \\
\midrule
gemini-2.5-flash-lite & 88.32\% & 36.86\% & 51.46\% & 6.17\% & 5.51\% \\
gemini-2.5-flash & 90.62\% & 64.29\% & 26.33\% & 5.39\% & 3.99\% \\
gemini-3.1-flash-lite-preview & 86.26\% & 58.49\% & 27.77\% & 7.82\% & 5.92\% \\
gemini-3-flash-preview & 85.73\% & 63.51\% & 22.21\% & 6.66\% & 7.61\% \\
claude-haiku-4.5 & 93.95\% & 45.04\% & 48.91\% & 3.21\% & 2.84\% \\
claude-sonnet-4.5 & 95.10\% & 53.19\% & 41.92\% & 3.17\% & 1.73\% \\
gpt-5-nano & 89.26\% & 49.12\% & 40.15\% & 6.33\% & 4.40\% \\
gpt-4o-mini & 95.60\% & 15.22\% & 80.38\% & 2.55\% & 1.85\% \\
qwen-2.5-VL-32B & 94.69\% & 42.29\% & 52.41\% & 2.88\% & 2.43\% \\
qwen-2.5-VL-7B & 94.36\% & 35.21\% & 59.15\% & 3.37\% & 2.26\% \\
qwen-2.5-VL-3B & 40.23\% & 15.34\% & 24.89\% & 15.96\% & 43.81\% \\
\bottomrule
\end{tabular}
}
\vspace*{-0.4cm}
\label{table: different LLM judge and rule-based judge}
\end{table}

\section{Discussion}
\label{section: Discussion}

This section presents a case study, discusses limitations, and outlines implications.

\subsection{Case Study}
\label{subsec: Case Study}

To further understand the limitations of VLMs in software architecture diagram understanding, we conduct a case study of incorrect model predictions.
By sampling 200 failure cases and manually analyzing their underlying causes, we identify several recurring error patterns related to visual reasoning and relation interpretation.
Specifically, \textit{Long Arrow}, \textit{Multi Arrow}, \textit{Non-Rightward Arrow}, and \textit{Overlap Arrow} occur in 24.0\%, 25.0\%, 32.0\%, and 33.0\% of the sampled failures, respectively.
Motivated by the prevalence of these patterns, we manually construct a separate hard set, distinct from SADU, containing 15 cases for each category.
\jie{these patterns came from nowhere}

% Specifically, the hard set focuses on four types of cases that frequently lead to incorrect predictions: \textit{Long Arrow}, \textit{Multi Arrow}, \textit{Non-Rightward Arrow}, and \textit{Overlap Arrow}.

% To further understand the limitations of VLMs in software architecture diagram understanding, we conduct a case study of incorrect model predictions.
% By examining representative failure cases, we identify several recurring error patterns related to visual reasoning and relation interpretation.
% Based on these observations, we manually construct a small hard set (15 cases each for 4 categories) that isolates these challenging scenarios.
% Specifically, we design four types of cases that frequently lead to incorrect predictions: \textit{Long Arrow}, \textit{Multi Arrow}, \textit{Non-Rightward Arrow}, and \textit{Overlap Arrow}

% To better understand the limitations of VLMs in software architecture diagram understanding, we conduct a case study of incorrect model predictions.
% We manually examine the errors made by the evaluated models and identify several recurring failure patterns related to visual reasoning and relation interpretation.
% Among all analyzed failures, \textit{Long Arrow}, \textit{Multi Arrow}, \textit{Non-Rightward Arrow}, and \textit{Overlap Arrow} account for approximately XX\%, XX\%, XX\%, and XX\% of the cases, respectively.
% Motivated by the prevalence of these patterns, we manually construct a small hard set to isolate them, with 15 cases for each category.

\begin{table}[h!]
\vspace*{-0.2cm}
\caption{Case study on hard cases}
\vspace*{-0.3cm}
\centering
\resizebox{\columnwidth}{!}{%
\begin{tabular}{l r r r r}
\toprule

Model & Long Arrow & Multi Arrow & Non-Rightward Arrow & Overlap Arrow \\
\midrule
gemini-2.5-flash-lite & 60.00\% & 20.00\% & 26.67\% & 20.00\% \\
gemini-2.5-flash & 86.67\% & 53.33\% & 33.33\% & 6.67\% \\
gemini-3.1-flash-lite-preview & 93.33\% & 86.67\% & 20.00\% & 33.33\% \\
gemini-3-flash-preview & 86.67\% & 40.00\% & 13.33\% & 20.00\% \\
claude-haiku-4.5 & 13.33\% & 20.00\% & 40.00\% & 26.67\% \\
claude-sonnet-4.5 & 40.00\% & 33.33\% & 33.33\% & 13.33\% \\
gpt-5-nano & 46.67\% & 13.33\% & 20.00\% & 6.67\% \\
gpt-4o-mini & 40.00\% & 13.33\% & 0.00\% & 0.00\% \\
qwen-2.5-VL-32B & 13.33\% & 20.00\% & 26.67\% & 0.00\% \\
qwen-2.5-VL-7B & 13.33\% & 6.67\% & 33.33\% & 6.67\% \\
qwen-2.5-VL-3B & 0.00\% & 26.67\% & 13.33\% & 0.00\% \\

\bottomrule
\end{tabular}
}
\vspace*{-0.25cm}
\label{table: hard benchmark}
\end{table}

Table~\ref{table: hard benchmark} reports the performance of different VLMs on this hard benchmark.
\textit{Long Arrow} appears to be the easiest for stronger models.
For example, \textit{gemini-3.1-flash-lite-preview} achieves 93.33\% accuracy.
However, this pattern is still challenging for other models, including \textit{claude-haiku-4.5} (13.33\%), \textit{qwen-2.5-VL-32B} (13.33\%), and \textit{qwen-2.5-VL-3B} (0.00\%), suggesting that tracing long-distance relations remains difficult.
\textit{Multi Arrow} cases also show a large performance gap across models.
\textit{gemini-3.1-flash-lite-preview} performs best at 86.67\%, 
while most other models remain below 35\%.
This result indicates that distinguishing among multiple nearby arrows is a particularly demanding form of visual relation reasoning.

\textit{Non-Rightward Arrow} and \textit{Overlap Arrow} are difficult for nearly all models.
For \textit{Non-Rightward Arrow}, the best result is only 40.00\%, achieved by \textit{claude-haiku-4.5}.
This suggests that VLMs are highly sensitive to deviations from standard horizontal or vertical arrow layouts.
For \textit{Overlap Arrow}, performance is even lower overall.
The best model, \textit{gemini-3.1-flash-lite-preview}, reaches only 33.33\%.
These results indicate that overlapping connectors are one of the most challenging visual patterns in our case study benchmark.

Overall, the case study confirms that current VLMs remain brittle when relation arrows become visually complex.
Even the best-performing models show clear degradation on non-standard arrow layouts, especially when arrows overlap or deviate from common orientations.
This finding suggests that software architecture diagram understanding is not only a matter of recognizing entities, but also of robustly parsing visually ambiguous relational structures.

\subsection{Limitations}

\textbf{Lack of Fine-grained Spatial Annotations.}
Our JSON representation captures semantic structure, including entities, relations, clusters, and labels, but does not include explicit positional or geometric information such as anchor points or relative layout constraints.
As a result, we cannot directly quantify geometry-dependent failure modes, such as errors caused by long arrows spanning large portions of a diagram, ambiguous edge crossings, or mistaken associations induced by proximity and visual clutter.
This also limits analyses that correlate model errors with layout complexity, such as edge length distribution, overlap rate, or crossing count.

\textbf{Data Selection and Contamination.}
Diagrams in SADU are collected from publicly available sources, such as Azure architecture documentation and common diagram-design platforms.
Because VLMs are often trained on large-scale web data, similar diagrams or related textual descriptions may have appeared in their training corpora.
However, we believe the practical impact of contamination is limited for several reasons.
First, SADU includes diagrams from multiple sources and covers diverse structures and layouts, which reduces the likelihood that models have memorized exact instances.
Second, the evaluation focuses on structured reasoning tasks, such as counting elements, identifying relations, and retrieving specific components, which depend on the precise content of each diagram rather than general architectural familiarity alone.
Third, the moderate performance of even the strongest models suggests that the benchmark remains nontrivial even if partial exposure exists.

\subsection{Implications}

% For \textbf{researchers}, this work highlights an important yet underexplored limitation of current VLMs.
% Although recent models achieve strong results on general multimodal tasks, our findings show that they remain weak on software architecture diagram understanding, particularly when tasks require structured reasoning over diagram elements.
% These results suggest that diagram understanding is qualitatively different from conventional visual question answering and may require more structure-aware modeling techniques.

For \textbf{researchers}, this work highlights an important but still underexplored limitation of VLMs.
Although VLMs perform strongly on general multimodal benchmarks, our results show that they remain inadequate for software architecture diagram understanding, especially when tasks require reasoning over diagram components. 
This suggests that software architecture diagram understanding is different from conventional visual question answering and may require models that more effectively capture structural semantics and spatial relationships, and domain-specific abstractions.

For \textbf{developers}, this work takes an initial step toward enabling intelligent tools that can more effectively interpret software design documentation.
Software architecture diagrams are central to the early stages of the software development lifecycle, where they communicate system structure, component interactions, and design intent.
However, our results show that even state-of-the-art VLMs achieve only moderate performance on this benchmark, indicating that they are not yet dependable for high-stakes development scenarios that require accurate architectural interpretation.
By exposing these limitations and providing a benchmark for measuring progress, SADU establishes a foundation for future AI systems that can more reliably integrate architectural knowledge into software engineering workflows.

\section{Related Work}
\label{section: Related Work}

\subsection{Benchmarks for Image Reasoning}

\textbf{General Multimodal Reasoning.}
MMMU~\cite{yue2024mmmu} and MMMU-Pro~\cite{yue2025mmmu} evaluate multimodal models on college-level subject knowledge and deliberate reasoning with highly heterogeneous visual inputs.
MMMU-Pro further strengthens the setup by filtering text-only-solvable questions, augmenting options, and embedding questions into images to enforce joint seeing-and-reading~\cite{yue2025mmmu}.
% Recent frontier model reports increasingly treat MMMU-Pro as a key headline metric for multimodal reasoning; for example, Gemini 3 Pro reports 81\% on MMMU-Pro, and GPT-5.4 reports 81.2\% (no tools) / 82.1\% (with tools) on MMMU-Pro~\cite{google2025gemini3, openai2026gpt54}.
However, strong MMMU-Pro scores do not imply robust multi-step reasoning in heterogeneous visual environments. 
OSWorld~\cite{xie2024osworld} shows a large gap to humans on real screenshot-based computer tasks due to GUI grounding and operational knowledge.
WildVision-Arena~\cite{lu2024wildvision} provides complementary signals of open-ended multimodal usefulness, but should not be conflated with targeted spatial-logic robustness.

\textbf{Scientific and Document Reasoning.}
For scientific charts, CharXiv constructs a realistic evaluation suite from 2,323 natural charts in arXiv papers and demonstrates sharp robustness drops under small chart/question perturbations, revealing a large gap to human performance~\cite{wang2024charxiv}.
Independent analysis further characterizes chart understanding as bottlenecked by visual perception and information extraction in current VLMs~\cite{liu2025perception}.
Chart2Code pushes beyond recognition to chart-to-code generation and editing, showing that even state-of-the-art models remain far from reliable performance on complex editing tasks~\cite{tang2025charts}.
For real-world documents, UniKIE-Bench provides a unified end-to-end KIE benchmark and highlights persistent challenges in grounding accuracy and layout-aware reasoning under complex layouts and long-tail fields~\cite{ji2026unikie}.

\subsection{Software Diagram Understanding}

The software engineering community has been moving from manual traceability and documentation upkeep toward agent-driven architecture knowledge management. 
Prior work highlights that architects often lack effective automation for keeping documentation current and checking conformance, resulting in ad-hoc, manual workflows \cite{ivers2025will}. 
This lack of rigor is a documented bottleneck; empirical surveys by \cite{nugroho2008survey} and \cite{lange2006effects} indicate that even when UML is used, a lack of synchronization between models and code often leads to high defect rates and documentation decay.
DRAFT-ing \cite{dhar2025draft} combines retrieval-augmented generation and fine-tuning to draft architecture decision records from distributed project artifacts.

However, many architectural artifacts remain visually formatted and difficult to integrate into automated pipelines. 
Industrial surveys report that diagramming tools are a dominant medium for communicating software architecture \cite{icepanel2024state}, while keeping documentation up to date and preventing architectural drift remain persistent challenges \cite{icepanel2025state}. 
This gap is exacerbated by the fact that design and architecture diagrams are frequently embedded as images in documents, hindering reuse and evolution \cite{chen2022automatically}. 
Motivated by the broader push from vision-as-parsing toward vision-centric reasoning evaluation \cite{man2025argus}, our benchmark focuses on downstream, high-level reasoning over component and deployment views rather than element extraction alone.

\section{Conclusion}
\label{section: Conclusion}

In this paper, we introduced SADU, a benchmark for evaluating VLMs on software architecture diagram understanding.
SADU covers diverse software architecture diagram families and provides a structured evaluation setting for probing both counting and retrieval reasoning over diagram elements.
Through experiments on 11 state-of-the-art VLMs, we found that software architecture diagram understanding remains challenging for current models.
Even the best-performing model achieves only moderate performance, and our analyses further show that VLMs still struggle with several important aspects of diagram understanding, including structured counting, precise retrieval, relational grounding, and visually challenging cases such as complex arrow layouts.
These findings suggest that software architecture diagram understanding remains an important challenge for AI-assisted software engineering.
% In future work, we plan to extend SADU with richer annotations, more diverse diagram types, and harder reasoning scenarios.
% We also hope this work can support the development of AI systems that better understand design documentation and integrate architectural knowledge more effectively into the software development lifecycle. \jingzhi{reference [18] is incomplete}

\section{Data Availability Statement}
\label{section: Data Availability Statement}
We release the benchmark, code, and evaluation results on our website https://doi.org/10.5281/zenodo.19339991
\newpage

\bibliographystyle{plain}
\bibliography{reference}

\end{document}